# Persistent spin helix on a wurtzite ZnO(10-10) surface: First-principles density-functional study

Moh Adhib Ulil Absor[1,2], Fumiyuki Ishii[3], Hiroki Kotaka[1,4], and Mineo Saito[3]

[1]Graduate School of Natural Science and Technology, Kanazawa University, Kanazawa 920-1192, Japan
[2]Department of Physics Universitas Gadjah Mada BLS 21 Yogyakarta Indonesia
[3]Faculty of Mathematics and Physics, Institute of Science and Engineering, Kanazawa University, Kanazawa 920-1192, Japan
[4]ISIR-SANKEN, Osaka University, 8-1 Mihogaoka, Ibaraki, Osaka, 567-0047, Japan

The persistent spin helix (PSH) that has been widely and exclusively studied in zinc-blende structures is revealed for the first time on the surface of a wurtzite structure. Through first-principles calculations of the ZnO(10-10) surface, a quasi-one dimensional orientation of the spin textures is identified. Further, the wavelength of this particular PSH is smaller than that observed with various zinc-blende quantum well structures, thus indicating that wurtzite-structured surfaces are suitable for spintronics applications.

Spin-orbit coupled systems have attracted consider-able scientific interest over recent years, as they allow for the manipulation of electron spin.[1] This spin-orbit coupling (SOC) appears to play an important role in new fundamental phenomena such as current-induced spin polarization[2] and the spin Hall effect.[3] The electric tunability of SOC has also been achieved by using gated semiconductor heterostructures,[4] thereby opening a new gateway to applications ranging from spintronics to quantum computing. Examples of some of the various spintronics devices that have been studied include spin-field effect transistors,[5] spin filters,[6] and spin qubit gates.[7]

Energy-saving spintronics devices are believed to be implementable using persistent spin helix (PSH) materials, as these induce a greatly enhanced spin relaxation time.[8-15] Theoretical studies have predicted that such PSH materials can be produced using [001]-oriented quantum wells (QWs) in which the Rashba and Dresselhauss terms are equal, or by using [110]-oriented QWs that are only affected by the Dresselhauss effect.[8] In ei-ther case, the spin-orbit coupling is linearly dependent on the electron momentum in specific directions, and so a one-dimensional orientation of the spin textures is generated.[8] The PSH states have been observed experimentally for [001]-oriented GaAs/AlGaAs QWs[9, 10, 14] and InAlAs/InGaAs QWs,[11-13] as well as for [110] oriented GaAs/AlGaAs QWs that exhibit uni-directional out-of-plane spin directions.[15]

Thus far, PSH has only been widely studied in zinc-blende semiconductors. However, experimental observation of high-quality two-dimensional systems in wurtzite structured semiconductors[16, 17] has let to them become increasingly viewed as a promising candidate for spintronics applications.[18] As such, there is clearly great value in achieving PSH in these semiconductors. In this Letter, we demonstrate that PSH is possible using a wurtzite ZnO(10-10) surface. By performing first-principles density-functional theory (DFT) calculations of the ZnO(10-10) surface, a quasi-one-dimensional orientation of the spin textures is identified and clarified by using a simplified spin-orbit Hamiltonian. Finally, it is revealed that the wavelength of this PSH is smaller than that seen with various zinc-blende QW structures.

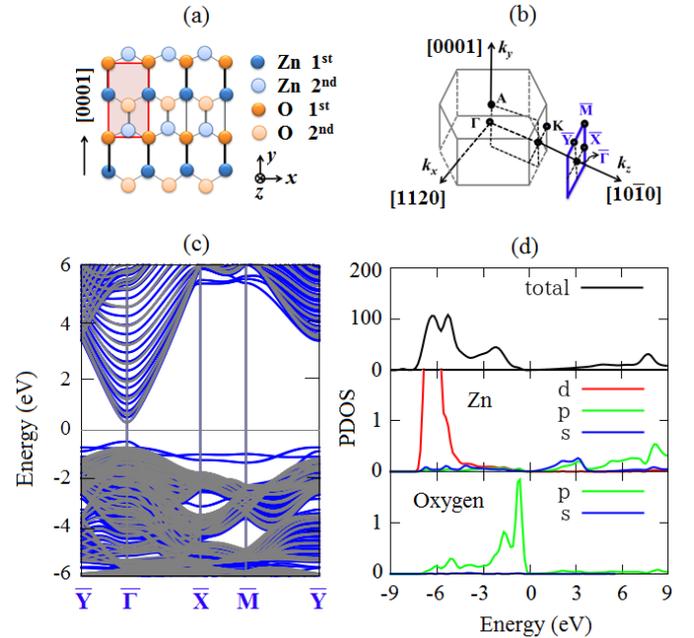

Fig. 1. (a) Top view of the geometric structure of ZnO(10-10) surface. The polar direction [0001] is set to be the $y$ direction. The unit cell of the slab is indicated by the red line. (b) Schematic view of the first Brillouin zone of the crystal and the surface Brillouin zone. (c) Band structures on the surface. The blue lines correspond to the band structures of ZnO(10-10) surface and the black lines correspond to those of the bulk system. (d) Total density of states and partial density of states projected to the first-layer atoms.

First-principles electronic-structure calculations based on the DFT were carried out within a generalized gradient approximation (GGA)[19] using the OpenMX code.[20] In a past study, it was found that the optimized lattice parameters of bulk ZnO are $a = 3.284$ Å and $c/a = 1.615$.[18] Surface calculations were carried out using a slab model consisting of 20-bilayers and terminated by hydrogen atoms on its backside [Fig. 1(a)]. The vacuum length was set to more than 15 Å in order to avoid inter-actions between neighboring slabs. Geometries were fully relaxed until the force acting on each atom was less than 1 meV/Å. Norm-conserving pseudopotentials[21] were used. The wave functions were expanded by a linear combination of multiple pseudo-atomic orbitals (LCPAOs) generated using a confinement scheme,[22,23] defined as Zn6.0-s²p²d², O5.0-s²p²d¹, and H5.0-s²p¹. For example, in the case of a Zn atom, Zn6.0-s²p²d² means that in this confinement scheme the cutoff radius is 6.0 Bohr[22, 23] and two primitive orbitals for the s, p, and d components are used. SOC was included in these fully-relativistic calculations, and the spin textures in $k$-space were calculated using the $k$-space spin density matrix of the spinor wave function.[24]

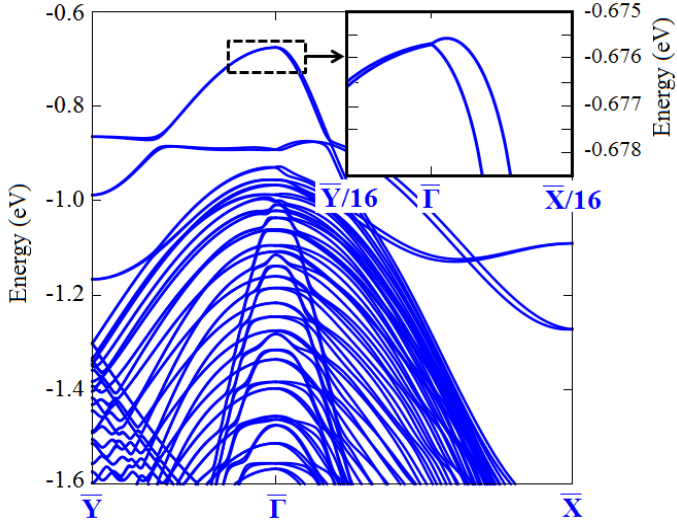

**Fig. 2.** Band structures near the VBM in the $Y$ - $\Gamma$ - $X$ direction. The insert shows the spin-split surface-state bands.

Calculations for a ZnO(10-10) surface revealed that the out-of-plane relaxation of the Zn and O atoms were $\delta_z$(Zn)=-0.28 Å and $\delta_z$(O)=-0.035 Å, respectively [Fig. 1(a)]. These values were slightly lower than the previous experimental values of $\delta_z$(Zn)=-0.45 Å and $\delta_z$(O)=-0.05 Å,[25] but are in a good agreement with past calculations [-0.036 to -0.21 Å ($\delta_z$(Zn)) and -0.04 Å ($\delta_z$(O))].[26-28] On the contrary, the in-plane relaxation of the Zn and O atoms, $\delta_y$(Zn)=0.19 Å and $\delta_y$(O)=-0.03 Å, respectively, were close to prior calculations [$\delta_y$(Zn)=0.116 Å and $\delta_y$(O)=-0.024 Å].[28] The occupied surface states were observed in band structures in the energy range of -1.3 to -0.65 eV [Fig. 1(b),(c)], from which the partial density of states (PDOS) was calculated through projection onto the surface atoms [Fig. 1(d)]. This revealed that the occupied surface state was characterized by O-2$p$ orbitals, which is consistent with the results of past calculations using a local density approximation (LDA).[27]

Given that the surface states are occupied, doping is expected to create a p-type system. Indeed, several researchers have succeeded in producing p-type, non-polar wurtzite ZnO films,[29, 30] giving credence to the notion of creating a p-type wurtzite ZnO(1010) surface.

The effect of SOC on the surface states can be seen in Fig. 2, wherein the band splitting is small in the $\Gamma$-$Y$ direction, but quite large in the $\Gamma$-$X$ direction. In these spin-split surface state bands, we find that the spin textures exhibit a quasi-one-dimensional orientation in the in-plane $y$ direction [Fig. 3(a)], yet also have out-of-plane

$z$ components [Fig. 3(b)]. These quasi-one-dimensional spin textures are expected to generate current in a direction perpendicular to the spin orientation and induce a greatly enhanced spin relaxation time through the PSH mechanism.[8] This is supported by the fact that a similar PSH has been observed in [110]-oriented zinc-blende QWs with out-of-plane spin orientations.[15]

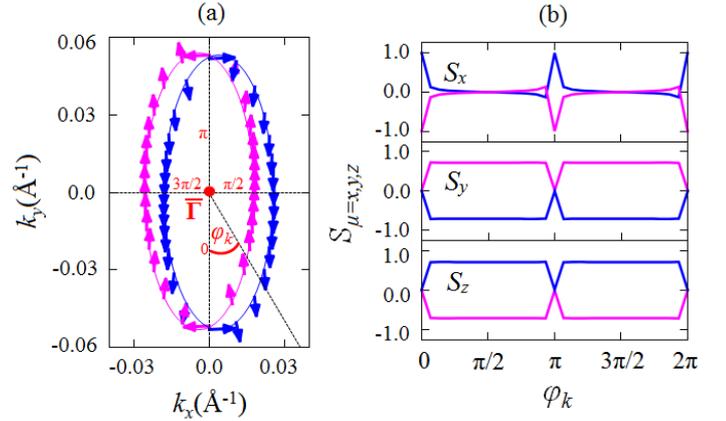

**Fig. 3.** (a) Spin textures of the surface states at the VBM. The band energy of the spin textures is 0.001 eV below the highest energy of the occupied surface state. The arrows represent the spin directions projected to the $k_x$-$k_y$ plane. (b) Relationship between rotation angle ($\varphi_k$) and spin components is shown.

To better understand the origin of spin textures, we must consider the SOC of surface states based on group theory.[31-34] This states that the ZnO(10-10) surface belongs to the symmetry point group $C_s$: the mirror reflections in operation in this symmetry transforms $(x,y,z)$ to $(-x,y,z)$. Given that a Hamiltonian SOC is completely symmetric in the $C_s$ point group and includes first-order terms over the wave vectors, the SOC can be expressed as, $H_{SOC} = \alpha_1 k_x \sigma_z + \alpha_2 k_x \sigma_y + \alpha_3 k_y \sigma_x$, where $k_x$ and $k_y$ are the wave vectors in the $x$- and $y$-directions, respectively, $\sigma_x$, $\sigma_y$, and $\sigma_z$ are Pauli matrixes, and $\alpha_1$, $\alpha_2$, and $\alpha_3$ are coupling constants that define the spin-orbit strength. Here, $\alpha_1$ is characterized by the in-plane electric field $E_y$, whereas $\alpha_2$ and $\alpha_3$ relate to the out-of-plane electric field $E_z$ that originates from the surface effect.

In the case of a bulk system oriented in the [10-10] direction, the out-of-plane electric field $E_z$ vanishes. Consequently, in the $k_x$-$k_y$ plane, both $\alpha_2$ and $\alpha_3$ are zero. This leads to a case where only the first term in the $H_{SOC}$ equation remains, meaning that the bands are spin degenerated in the $\Gamma$-$Y$ direction and the spin textures are oriented to the out-of-plane $z$-direction [See the supplementary materials]. In surface states, on the contrary, a band split is introduced in the $\Gamma$-$Y$ direction due to the third term in $H_{SOC}$ equation [Fig. 2]. Furthermore, as a result of the second term in this equation, a tilting of the spin textures in the in-plane $y$-direction is induced [Fig. 3]. It can therefore be concluded that the above spin-orbit Hamiltonian of the surface state matches well with the calculated results, i.e., the band split in the $\Gamma$-$Y$ direction and the tilt of the spin textures.

Since $H_{SOC}$ is strongly affected by the electric field as mentioned above, the origin of the spin textures can be further clarified by studying the electric polarization.[18] Given that the spin-split surface state is strongly localized in the first two-bilayers [Fig. 4(a)], a strong electric polarization is expected to occur in these bilayers. To clarify this, the layer-dependence of the electric polarization was calculated using a point charge model (PCM) for $Zn^{2+}$ and $O^{2-}$ ions in the bilayer in order to evaluate the polarization difference: $\Delta \mathbf{P} = \mathbf{P}(c/a,u) - \mathbf{P}(c/a, u_{ideal})$. Here, $c/=a$ and $u$ are the lattice constant ratio and internal parameter for a given optimized structure, respectively, and $u_{ideal} = 0.375$. As shown in Fig. 4(b), this reveals that the strongest electric polarization is identified near the first bilayer.

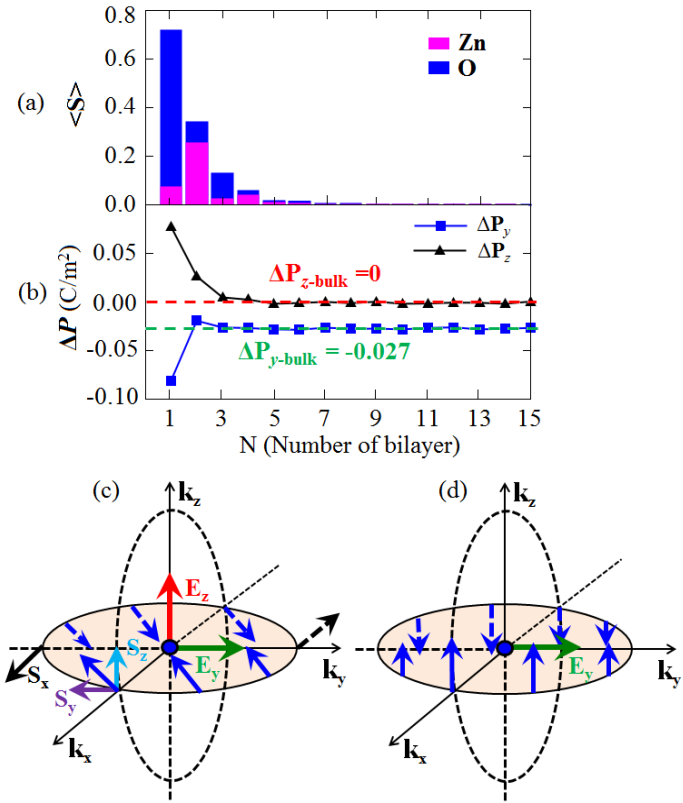

**Fig. 4.** (a) Expected values of spin projected to the atoms in each bilayers. The calculations are performed in the surface state in Fig. 3(a). The top of the surface is represented by N= 1. (b) Calculated data of the in-plane and out-of-plane electric polarizations ($\Delta P_y$, $\Delta P_z$) in each bilayers. The electric polarizations are calculated by using the PCM. Schematic view of the spin textures and electric fields for the case of the surface (c) and bulk systems (d).

Electric polarization in the out-of-plane $\Delta P_z$ and in-plane $\Delta P_y$ directions was calculated to be 0.077 C/m² and -0.081 C/m², respectively. These values indicate that the electric field in the out-of-plane $E_z$ direction was comparable to that in the in-plane $E_y$ direction, which would induce a tilting of the spin orientation [Fig. 4(c)]. However, only the in-plane electric polarization was observed in the case of bulk system [$\Delta P_{y\ bulk} = 0.027$ C/m²]. This led to the generation of the in-plane electric field $E_y$, but induced a spin orientation in a fully out-of-plane $z$-direction [Fig. 4(d)]. This confirmed that the spin textures in Fig. 3 were consistent with the PCM results.

PSH that induces a greatly enhanced spin relaxation time has recently been extensively studied,[8-15] with our calculations indicating that this is in fact achieved using the ZnO(10-10) surface. Since the spin textures in the calculated results show a quasi-one-dimensional orientation, a magnetic field is induced in a direction parallel to spin orientation. This inhibits the precession of the spins, thereby increasing the spin relaxation time. A similar mechanism behind long spin relaxation times has been reported in [110]-oriented zinc-blende QWs,[35-37] suggesting that the ZnO(10̄10) surface could provide an efficient spintronics device.

The spin-orbit strength of the PSH ($\alpha_{PSH}$), a variable of interest in spintronics device applications, was calculated using the band dispersion in Fig. 2. This revealed that the value of $\alpha_{PSH}$ was quite substantial (34.78 meVÅ), and much larger than those observed in the PSH of various zinc-blende n-type QW structures of GaAs/AlGaAs [(3.5 to 4.9 meVÅ),[10] (2.77 meVÅ)[14]] and InAlAs/InGaAs [(1.0 meVÅ)[12] and (2.0 meVÅ)[13]]. This large value of $\alpha_{PSH}$ should ensure a smaller wavelength of PSH ($\lambda_{PSH}$), which is important to the miniaturization of spintronics devices. The calculated value $\lambda_{PSH}$ (0.19 μm) was in fact one order less than that observed in the direct mapping of PSH (7.3 to 10 μm)[10] and the resonant inelastic light-scattering measurement (5.5 μm)[14] of GaAs/AlGaAs QWs.

In summary, by studying PSH on a ZnO(10-10) surface through first-principles DFT calculations, it has been shown that the spin textures exhibit a quasi-one-dimensional orientation. We also found that the wavelength of this PSH is smaller than those observed on various zinc-blende quantum well structures. These results indicate that PSH can be achieved using a wurtzite(10-10) surface or interface with in-plane electric polarization and mirror symmetry, rendering it suitable for spintronics applications.

**Acknowledgement** Part of this research was funded by the MEXT HPCI Strategic Program. This work was partly supported by Grants-in-Aid for Scienti c Research (Nos. 25390008, 25790007, 25104714, 26108708, and 15H01015) from the Japan Society for the Promotion of Science (JSPS). The computations in this research were performed using the supercomputers at the Institute for Solid State Physics (ISSP) at the University of Tokyo. One of the authors (M.A.U.A) thanks the Directorate General of Higher Education (DIKTI), Indonesia and Kanazawa University, Japan, for financial support through the Joint Scholarship Program.

### Supplemental information for "Persistent spin helix on a wurtzite ZnO (10-10) surface: First-principles density-functional study"

In this supplementary material, we present the result of first-principles DFT calculation on ZnO bulk system oriented on the [10-10] direction. Wurtzite ZnO forms a hexagonal close-packed lattice where the in-plane and axial lattice parameters are represented by $a$ and $c$, respectively [Fig. S1 (a)]. The conventional unit vectors are given by $\bar{a}_1 = (1/2, \sqrt{3}/2, 0)a$, $\bar{a}_2 = (1/2, -\sqrt{3}/2, 0)a$, and $\bar{a}_3 = (0, 0, c/a)a$ where $a$ and $c$ are the lattice constants in the $a$ and $c$-directions, respectively. The Zn atoms are located at $(0,0,0)$ and $(2/3, 1/3, 1/2)$ whereas the O atoms are located at $(0,0,u)$ and $(2/3, 1/3, u + 1/2)$. The length of Zn-O bond along the $c$-axis is given by $d = uc$.

Here, we introduce new unit vectors which are suitable for describing non-polar [10-10] surfaces. The new unit vectors are $\bar{a}_1 = (1,0,0)a$, $\bar{a}_2 = (0, c/a, 0)a$, and $\bar{a}_3 = (0, 0, \sqrt{3})a$ with eight atoms per unit cell. In the new unit vectors, the polar [0001] and non-polar [10-10] directions are set to be the $y$- and the $z$- directions, respectively [Fig. S1 (b)]. Our calculations of the optimized lattice constants show that $a$ = 3.2845 A, $c$ = 5.3029 Å, $c/a$ = 0.6151, and $u$ = 0.3791. These values are consistent with our previous results.[1]

Here, we study the effect of SOC on the [10-10]-oriented ZnO bulk system. We focus on the top of valence band maximum (VBM) along the high symmetry line in the first Brillouin zone [Fig. S1(c)]. As shown in Fig. 2, the bands are spin degenerated in the $\Gamma$-$Y$ direction, whereas they split in the $\Gamma$-$X$ direction. In this spin-split band, the fully out-of-plane orientations of the spin textures are found [Fig. 3].

Considering the fact that this system has the in-plane electric field in the $y$ directions, which originates from the polarity of the present system, the SOC can be expressed by $H_{SOC} = \alpha_1(k_x\sigma_z + k_z\sigma_x)$, where $k_x$ and $k_z$ are the wave vectors in the $x$- and $z$- directions, respectively, $\sigma_x$ and $\sigma_z$ are the $x$ and $y$ components of Pauli matrixes. In this expression, $\alpha_1$ is the Rashba spin-orbit strength, which is proportional to the in-plane electric filed $E_y$. In the case that $k_z = 0$, only the first term of $H_{SOC}$ remains. In this case, spin degenerated bands are induced in the $\Gamma$-$Y$ direction and they split in the $\Gamma$-$X$ direction [Fig. S2]. Furthermore, due to the first term of $H_{SOC}$, the fully out-of-plane directions of the spin textures are generated [Fig. S3].

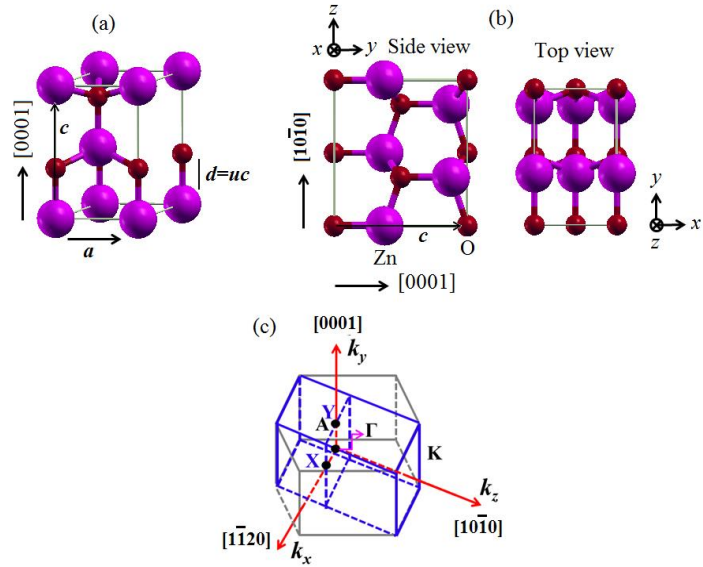

**Fig. S1**. Crystal structures of wurtzite bulk ZnO (a) and [1010]-oriented bulk ZnO (b) and first Brillouin zone (c). Black and blue lines represent the first Brillouin zone of the wurtzite bulk system and [1010]-oriented bulk system, respectively.

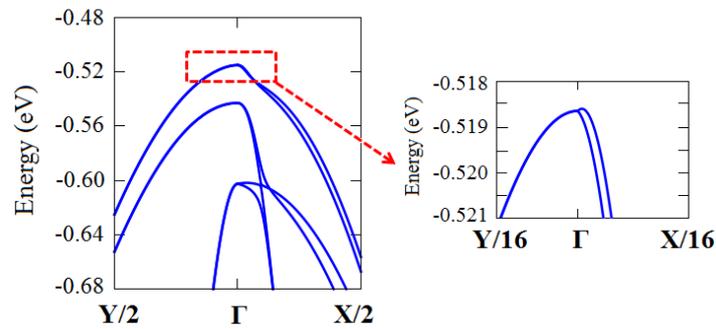

**Fig. S2**. Band structures near the VBM in the $Y$ - $\Gamma$ - $X$ direction. The insert shows the spin-splitt band at the top of VBM.

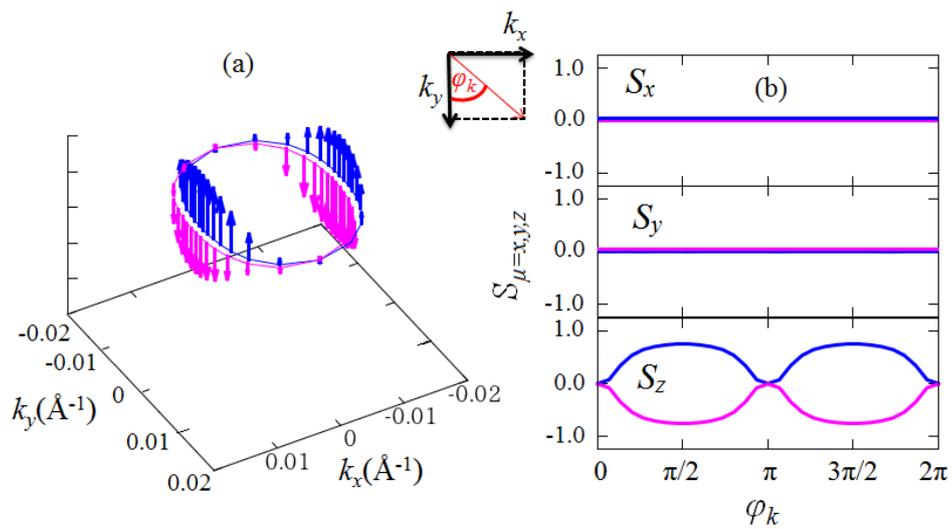

**Fig. S3**. (a) Spin textures of VBM. The band energy of the spin textures is 0.001 eV below the highest energy of the occupied states. The arrows represent the spin directions. The blue and pink lines correspond to iso-surfecae of the above-mentioned band energy. (b) Relationship between rotation angle ($\varphi_k$) and spin components.